\documentclass[journal]{IEEEtran}
\input{ds.def}

\newcommand{\AQLNGS}{INFN Laboratori Nazionali del Gran Sasso, Assergi (AQ) 67100, Italy}
\newcommand{\AQGSSI}{Gran Sasso Science Institute, L'Aquila 67100, Italy}

\newcommand{\Houston}{Department of Physics, University of Houston, Houston, TX 77204, USA}

\newcommand{\Princeton}{Physics Department, Princeton University, Princeton, NJ 08544, USA}

\usepackage{booktabs,caption}

\begin{document}
\title{Development of a very low-noise cryogenic pre-amplifier for large-area SiPM devices}
\author{
  	Marco~D'Incecco,
	Cristiano~Galbiati,
	Graham~K.~Giovanetti,
	George~Korga,
	Xinran Li,
	Andrea~Mandarano,
	Alessandro~Razeto,
	Davide~Sablone,
	and~Claudio~Savarese
	\thanks{Manuscript revisioned on \today.}
	\thanks{We acknowledge support from \NSF\ (US, Grant PHY-1314507 for Princeton University), the Istituto Nazionale di Fisica Nucleare (Italy) and Laboratori Nazionali del Gran Sasso (Italy) of INFN.}
	\thanks{Work at Princeton University was supported by Fermilab under Department of Energy contract DE-AC02-07CH11359.}
	\thanks{M.~D'Incecco is with \AQLNGS.}
	\thanks{C.~Galbiati, G.K.~Giovanetti and X.~Li are with \Princeton.}
	\thanks{G.~Korga is with \Houston\ and \AQLNGS.}
	\thanks{A.~Mandarano, and C.~Savarese are with \AQGSSI\ and \AQLNGS.}
	\thanks{D.~Sablone and A.~Razeto are with \AQLNGS\ and \Princeton.}
	\thanks{Corresponding Author: sarlabb7@lngs.infn.it}
}
%---
\maketitle
%---
\begin{abstract}
Silicon Photomultipliers (\SiPMs) are an excellent candidate for the development of large-area light sensors. Large \SiPM-based detectors require low-noise pre-amplifiers to maximize the signal coupling between the sensor and the readout electronics. This article reports on the development of a low-noise transimpedance amplifier sensitive to single-photon signals at cryogenic temperature. The amplifier is used  to readout a 1~cm$^2$ \SiPM\ with a signal to noise ratio in excess of 40.
\end{abstract}
%---
\IEEEpeerreviewmaketitle
%---
\section{Introduction}
\label{sec:intro}

The improvements in performance of \SiPMs\ at cryogenic temperature~\cite{Acerbi:2017gy} and the availability of low-noise hetero-junction amplifiers open the door to the development of cryogenic, large-area, \SiPM-based photo-detectors as replacements for photomultiplier tubes. We introduce and describe a low-noise transimpedance amplifier designed to instrument large-area assemblies of \SiPMs\ intended for operation at cryogenic temperature. As the scale and sensitivity of rare-event physics searches grows, so does the need for large, high efficiency light detectors. These amplifiers were designed for use in the \DSk\ dark matter detector, which will instrument an active volume of about \DSkActiveMass\ of liquid argon with \SI{14}{\square\meter} of \SiPMs~\cite{Aalseth:2017tr}. Due to their excellent signal to noise performance, they are also generally useful as the building block for realization of other large \SiPM\ assemblies.

Development of the amplifier was done in a dedicated cryogenic setup, consisting of a \LNGSCryoSetupCryocoolerModel\ pulse tube cryocooler mounted on the top flange of a stainless steel cryostat evacuated to \LNGSCryoSetupOperatingPressure. The cryocooler cold finger extends into the cryostat and holds the devices under test. The temperature of the cold finger is monitored by a set of platinum RTDs and regulated by a \LNGSCryoSetupTemperatureControllerModule\ to within \SI{1}{\kelvin} of a set-point temperature. The setup is capable of operating within the range from \LNGSCryoSetupTemperatureRange.

The core of the transimpedance amplifier is a LMH6629 high-speed operational amplifier from Texas Instruments. The cryogenic characterization of the LMH6629 will be described in Section~\ref{sec:6629}.  The characteristics of the \SiPMs\ used for testing, the design of the transimpedance amplifier and the analysis algorithms will be discussed in Section~\ref{sec:tia}. Finally, the results of the read-out of a \SI{1}{\square\cm} \SiPM\ will be described in Section~\ref{sec:results}.

\section{LMH6629 Characterization}
\label{sec:6629}
The performance of standard Silicon Bipolar Junction Transistors (BJT) degrades at cryogenic temperatures~\cite{Dumke:1981dy} whereas that of Hetero-junction Bipolar Transistors (HBT) does not, making HBTs an interesting alternative to CMOS technology for cryogenic applications.  An HBT is a type of bipolar junction transistor that uses different semiconductor materials for the emitter and base regions.  Among the available technologies, Silicon Germanium (SiGe) HBTs, which are manufactured by many foundries for high bandwidth applications, are particularly well-suited for cryogenic operation because of their bandgap-engineered base~\cite{Cressler:1994ck}.

The LMH6629 is a high speed, ultra-low noise amplifier fabricated with SiGe technology by Texas Instruments. The device is designed for applications requiring a wide bandwidth with high gain and low noise~\cite{TexasInstruments:2016wl}. The amplifier is available in two packages, a standard SOT-23-5 and a lead-less WSON-8. The WSON-8 package has selectable internal compensation for operation at low gains. Without this additional compensation enabled, the amplifier is stable only for gains larger than \SI{10}{V/V} at cryogenic temperatures.  This work focuses on the WSON-8 package, which was selected because its die attach pad improves thermalization.

Measurements of the LMH6629 were performed on custom FR4 printed circuit boards installed on the cold finger of the cryogenic setup. The passive components in the circuit were carefully selected for their cryogenic compatibility. All of the capacitors in the circuit are either C0G/NP0 or PPS~\cite{Teyssandier:2010tw}, and the resistors are based on thin or thick metal film.

\begin{figure}[!t]
  \centering
  \includegraphics[width=0.63\columnwidth]{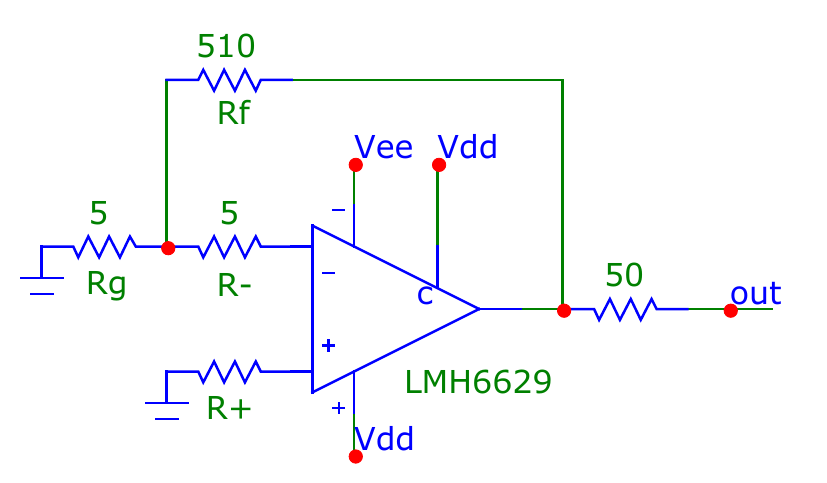}
  \caption{Schematic of the circuit used to measure the intrinsic noise sources of the LMH6629.} % with the internal compensation reduced.}
  \label{fig:noise-schem}
%\end{figure}
%\begin{figure}[!t]
  \vskip 5mm
  \centering
  \includegraphics[width=\columnwidth]{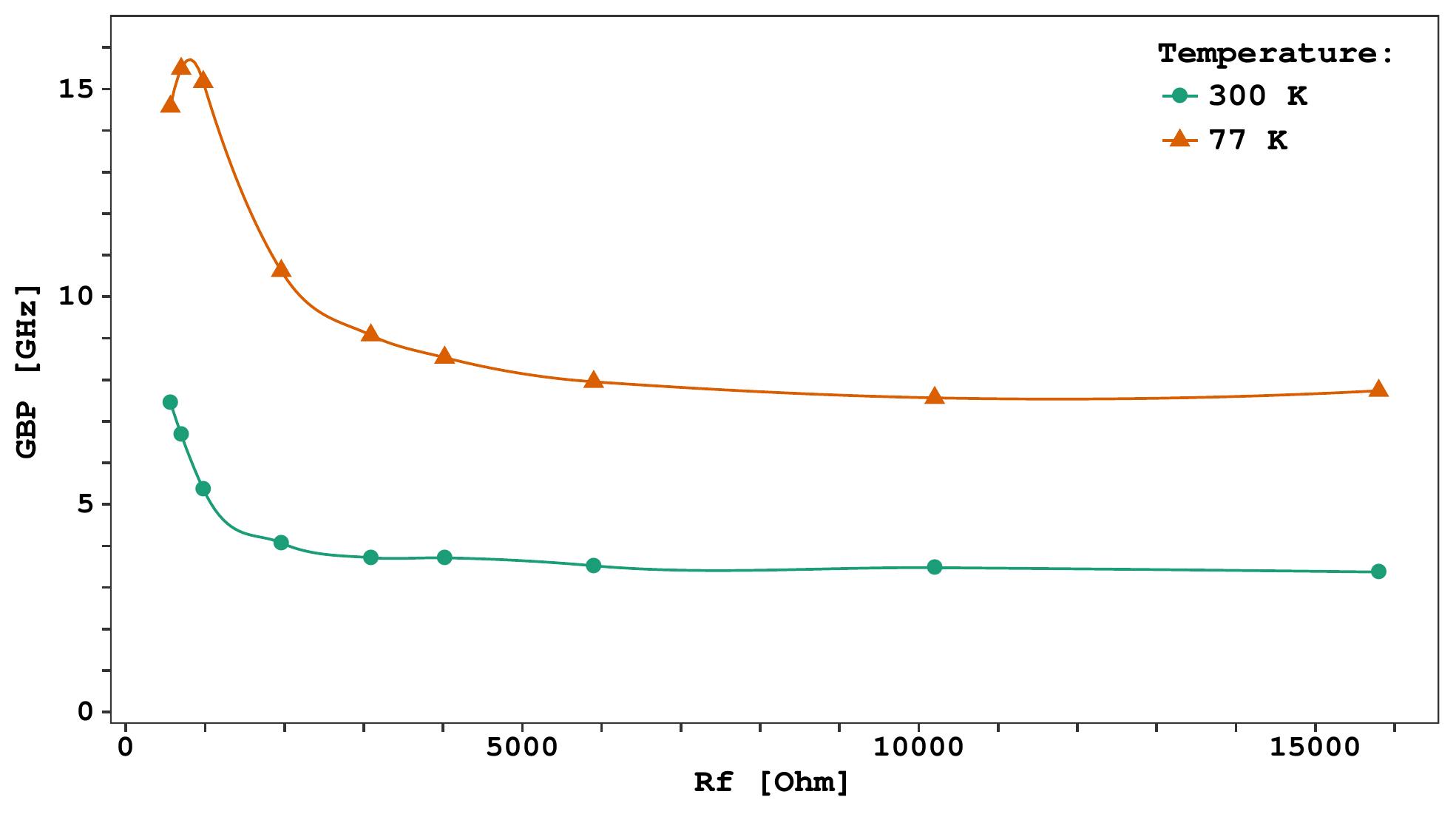}
  \caption{Gain bandwidth product of the LMH6629 test circuit in the non-inverting configuration versus feedback resistor value ($R_g = \SI{50}{\ohm}$, \SI{5}{V}). The LMH6629 is operated with the reduced internal compensation enabled. The lines are drawn to guide the eye.}
  \label{fig:gbp-rf}
\end{figure}

\begin{figure}[!t]
  \centering
  \includegraphics[width=\columnwidth]{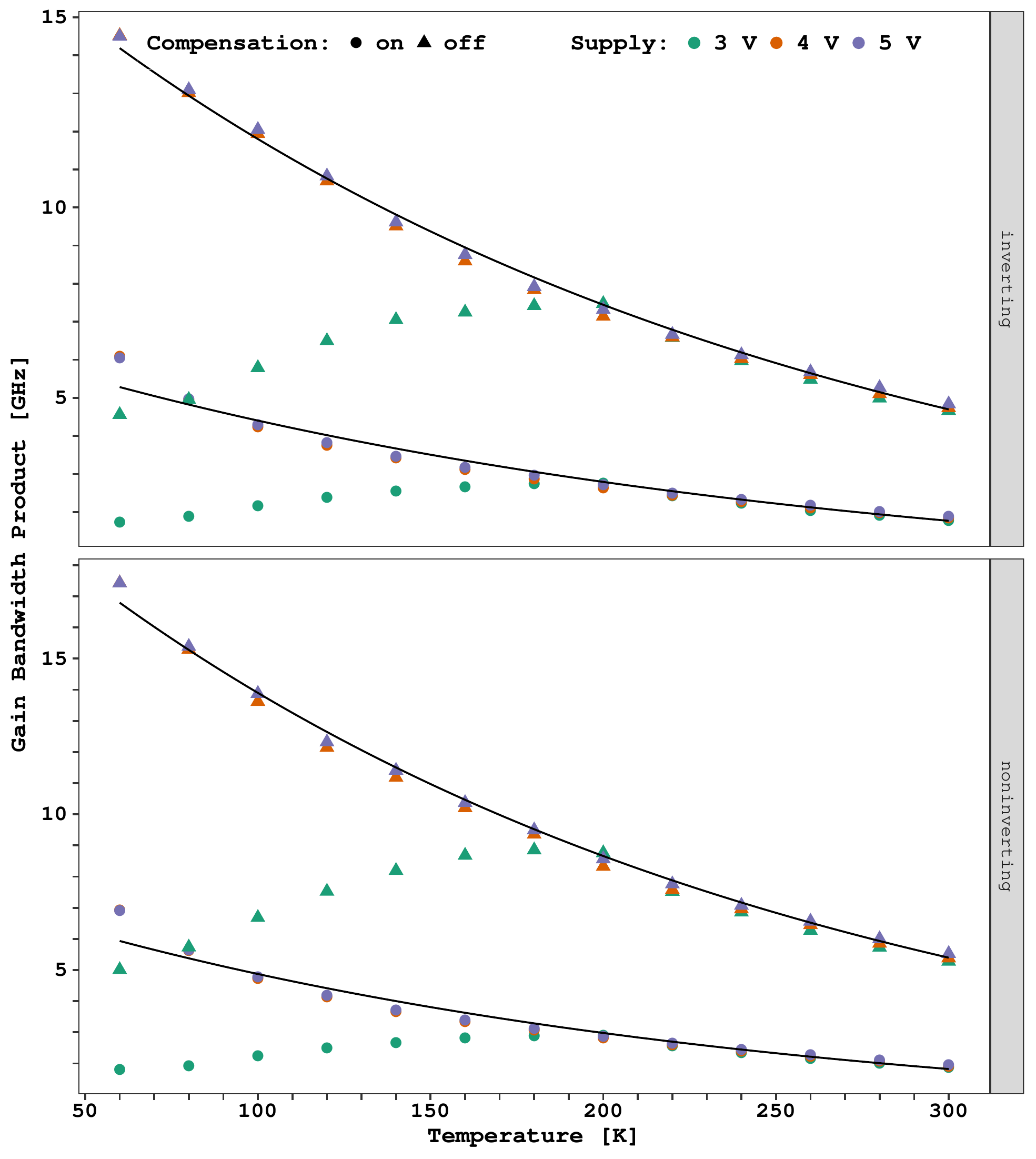}
  \caption{Gain bandwidth product of the LMH6629 circuit with $R_f=\SI{1}{\kohm}$ in inverting and non-inverting configurations versus temperature. The black lines correspond to exponential fits.}
  \label{fig:gbp}
\end{figure}

\subsection{Bandwidth}
\label{sec:bandwidth}
The measurement of the intrinsic bandwidth of the LMH6629 is complicated by parasitic effects from the circuit in which the chip is operated. In fact, a small capacitance on the feedback path can significantly reduce the measured bandwidth. Such behavior is demonstrated in Figure~\ref{fig:gbp-rf}, where the Gain Bandwidth Product (GBP) of the test circuit is shown for different feedback resistor ($R_f$) values. The GPB rapidly decreases as $R_f$ increases, eventually reaching a plateau. The spice simulation described in Section~\ref{sec:simulation} models this effect by including a \SI{0.2}{pF} feedback parasitic capacitance ($C_f$).

The GBP was measured by extracting the noise gain and the frequency of the half-power point ($F_{-3dB}$) from spectra collected with an Agilent model 5071c Vector Network Analyzer (VNA) configured for S$_{21}$ scans. Figure~\ref{fig:gbp} shows the GBP temperature dependence measured in both the inverting and non-inverting configurations ($R_f = \SI{1}{\kohm}$).  The GBP of the amplifier is exponential with temperature as shown by the fit in Figure~\ref{fig:gbp}. When operated with a \SI{3}{\volt} power supply, the amplifier performance quickly degrades below \SI{160}{\kelvin}.

\subsection{Input bias current}
\label{sec:bias}
The input bias current ($i_b$) of the LMH6629 was determined using the circuit shown in Figure~\ref{fig:noise-schem} with $R_+=\SI{620}{\ohm}$. The DC output of the circuit is given by $i_b\, (R_+-(R_-+R_g\parallel{R_f}))\, G$, with $G = R_f/R_g + 1$.  The results of the measurement are shown in Figure~\ref{fig:offset}. At room temperature, $i_b=\SI{-15}{\mu A}$, in agreement with the value reported on the datasheet. It is interesting to note that the input bias current rapidly decreases at low temperature, reaching a plateau of about \SI{\pm 1}{\micro\ampere} at about \SI{150}{\kelvin}.

\subsection{Input noise}
\label{sec:noise}
\begin{figure}[!t]
  \centering
  \includegraphics[width=\columnwidth]{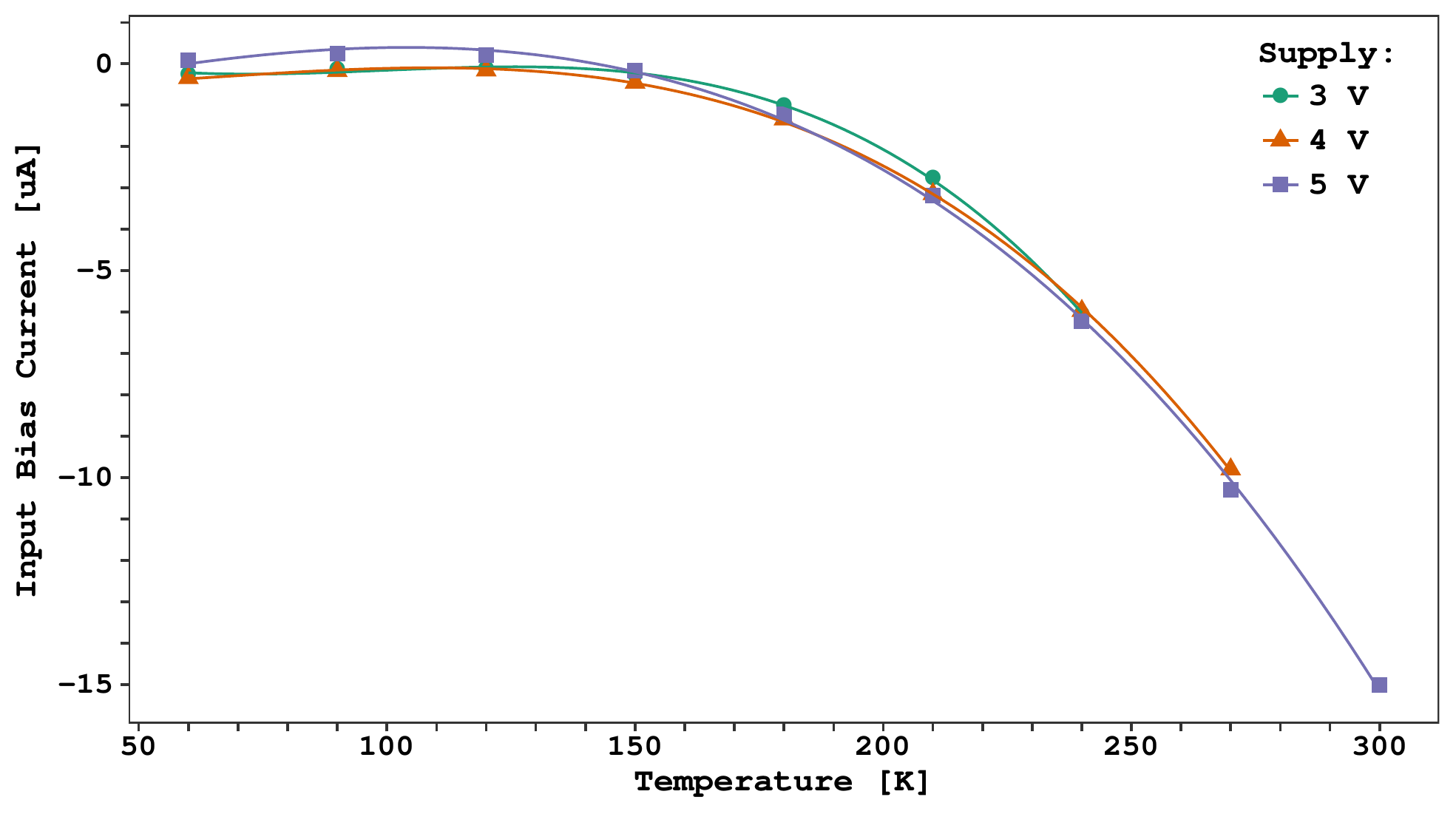}
  \caption{Input bias current of the LMH6629 versus temperature with different power supply voltages. The lines are drawn to guide the eye.} %. Below \SI{150}{\kelvin}, the input bias current is negative when using a \SI{5}{\volt} supply.} 
  \label{fig:offset}
%\end{figure}
%\begin{figure}[!t]
  \vskip 3mm
  \centering
  \includegraphics[width=\columnwidth]{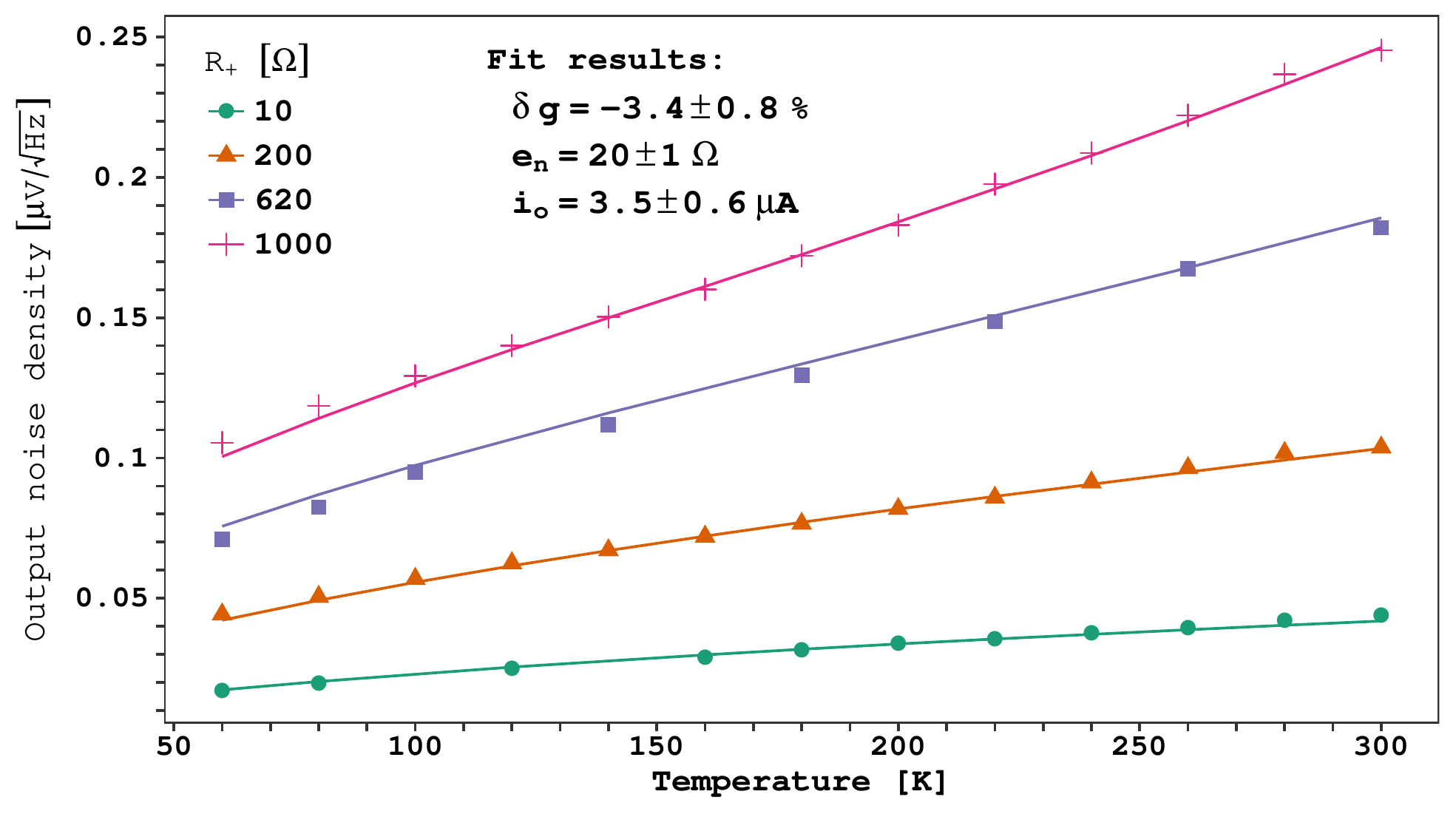}
  \caption{Output noise density at \SI{1}{\MHz} of the circuit shown in Figure~\ref{fig:noise-schem} operated at \SI{4}{\volt} versus temperature. The solid line represents the best fit to the data of the model described in the text.}
  %, which accounts for the Johnson noise of the resistors in the circuit and the intrinsic noise sources of the LMH6629. The fit describes the measurements at better than 3.4\%.}
  \label{fig:noise}
\end{figure}

Measurements of the voltage ($e_n$) and current ($i_n$) noise terms as a function of temperature were done with the LMH6629 configured as shown in Figure~\ref{fig:noise-schem}. The amplifier output was acquired with an R\&S FSV~7 spectrum analyzer and an R\&S RTO1024 oscilloscope operating as a spectrum analyzer. The output noise density was determined at \SI{1}{\MHz} in a flat region of the power spectra, far from any environmental noise pickup and the half-power point. Figure~\ref{fig:noise} shows the results for several values of $R_+$ along with fits to the noise model described below.

The output noise density ($N_o$) as a function of temperature ($T$) can be expressed as
\begin{equation}
  N_o = \frac{G}{2} \sqrt{4 k_B T R^J_{eq} + e_n^2(T) + i_n^2(T) R_i^2}
  \label{eq:noise}
\end{equation}
where $k_B$ is the Boltzmann constant,
\begin{gather*}
  R^J_{eq} = R_+ + R_-^* \xi^2, ~ R^2_i = R_+^2 + R_-^{*2} \xi^2, \\
  \xi = (G-1)/G \textrm{, and} ~ R_-^* = R_f\parallel{R_g} + R_-.
%  &G  = R_f/R_g + 1 ~ \textrm{and} ~ \xi = (G-1)/G, \\
%  &R^J_{eq} = R_+ + \left(R_f\parallel{R_g} + R_- \right) \xi^2, \\
%  &\textrm{and}~R^2_i = R_+^2 +  \left(R_f\parallel{R_g} + R_-\right)^2 \xi^2.
\end{gather*}
Equation~\ref{eq:noise} accounts for the Johnson-Nyquist noise of the resistors in the circuit and for the intrinsic noise sources of the LMH6629.

The fit model assumes that $e_n(T)$ behaves like a Johnson-Nyquist noise source with resistance $R_n$ ($e_n(T) = \sqrt{4 k_B T R_n}$) and that $i_n(T)$ is due to the shot noise of the input current ($i_n = \sqrt{2 e^- \left(|i_b(t)|+|i_o|\right)}$ with the input bias current, $i_b(T)$, measured in Section~\ref{sec:bias} and $i_o$ assumed constant). The best fit values are:
\begin{compactitem}
\item $R_n = \SI{25\pm1}\ohm$, which at \SI{300}{\kelvin} is equivalent to \SI{0.64\pm0.01}{\nV/\sqrt{Hz}}, compared to \SI{0.69}{\nV/\sqrt{Hz}} on the datasheet;
\item $i_o = \SI{3.5\pm0,6}{\mu A}$, for a total shot noise density of \SI{2.4\pm0.1}{\pA/\sqrt{Hz}}, compared to \SI{2.6}{\pA/\sqrt{Hz}} on the datasheet;
\item and $\delta G = -3.4\pm0.8~\%$, where $\delta G$ is the relative variation of the gain with respect to the theoretical value and is allowed to float during the fit.
\end{compactitem}
Overall, the experimental values are well described by the model. The standard deviation of the relative residuals is \SI{2.5}{\percent}, mostly due to the measurements performed with large $R_+$ values. This indicates that the accuracy of the scaling model of the current noise with temperature needs to be improved. However, in this work $i_n$ has no effect on the overall noise budget given the small value of the resistors in use (see Section~\ref{sec:pre:performances}).

\subsection{Dynamic range}
\label{sec:dyn}
\begin{figure}[!t]
  \centering
  \includegraphics[width=\columnwidth]{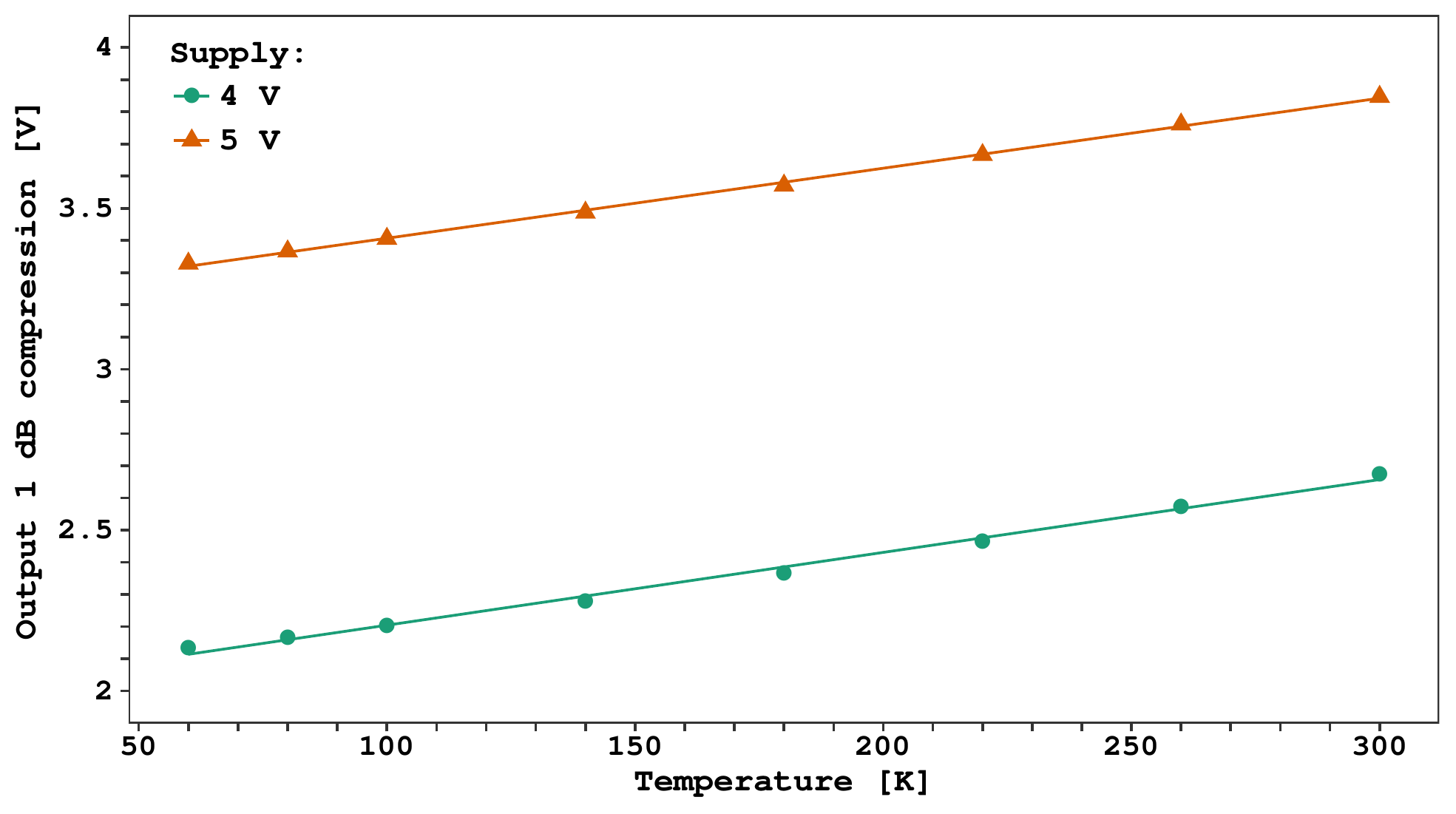}
  \caption{Peak to peak output compression (1 dB) of the LMH6629 versus temperature. The lines represent a linear fit to the data.}
  \label{fig:compression}
  \vskip 3mm
%\end{figure}
%\begin{figure}[!t]
  \centering
  \includegraphics[width=\columnwidth]{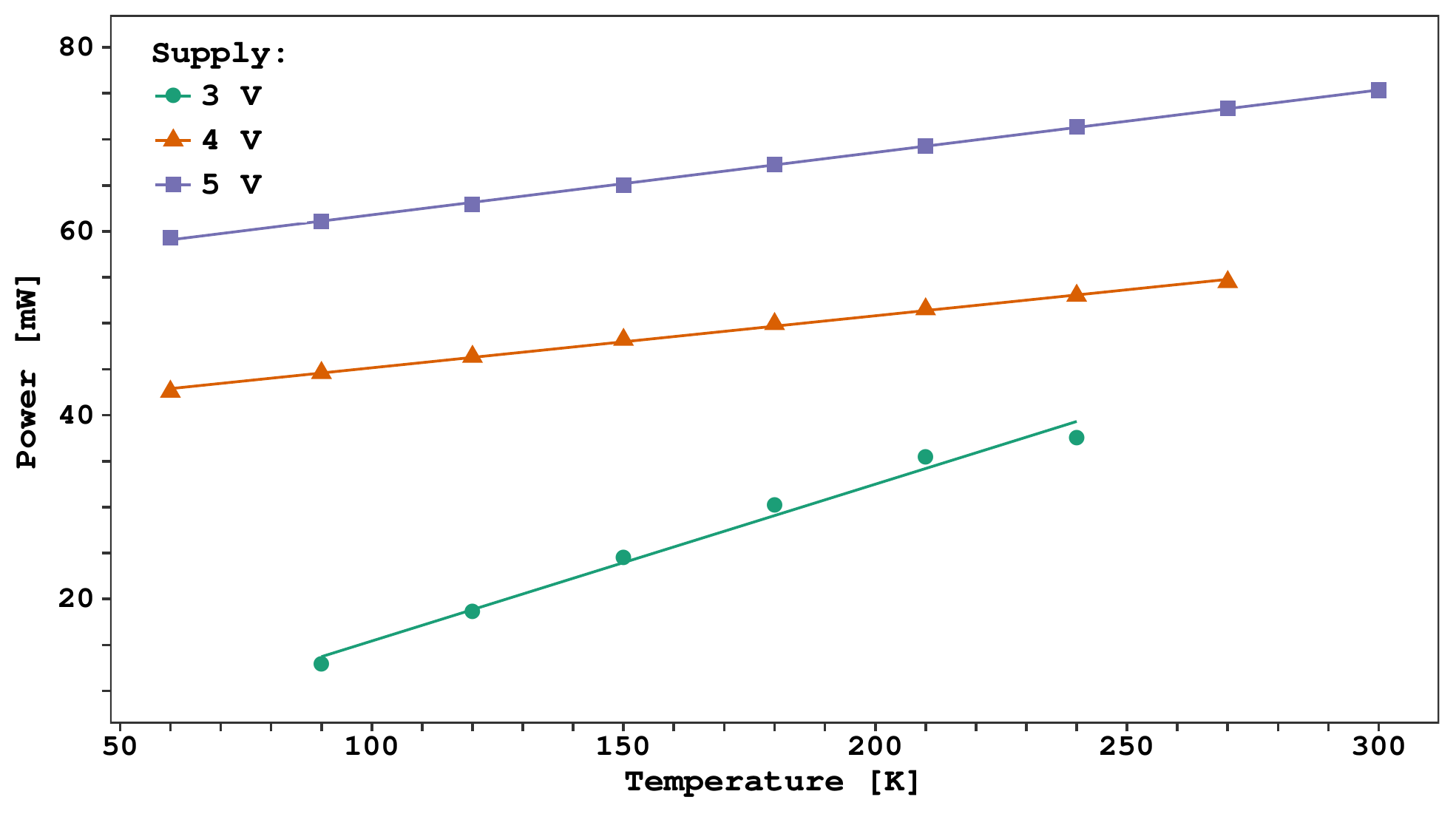}
  \caption{Power consumption of the LMH6629 versus temperature. The lines represent a linear fit to the data.}
  \label{fig:power}
\end{figure}
The dynamic range of the LMH6629 output was measured using the VNA configured for a power sweep at \SI{1}{\MHz} in S$_{21}$ mode. Figure~\ref{fig:compression} shows the \SI{1}{\dB} compression point as a function of temperature for two different power supply voltages (\SI{4}{\volt} and \SI{5}{\volt}) along with a linear fit to the data. Figure~\ref{fig:compression} plots the peak to peak voltage. When the circuit is used as a pulse amplifier, the effective dynamic range is halved.

\subsection{Power consumption}
\label{sec:power}
The power consumption of the LMH6629 was estimated by measuring the current flowing in the circuit when no load was connected to the output. Figure~\ref{fig:power} shows the results for various supply voltages along with a linear fit to the data.

\subsection{Simulation}
\label{sec:simulation}
%\begin{figure}[!t]
%  \centering
%  \includegraphics[width=\columnwidth]{figures/gbp_mc-vs-T.pdf}
%  \caption{SPICE simulation of the GBP with a \SI{5}{\volt} supply voltage, the internal compensation disabled, and with a \SI{0.2}{\pF} capacitance on the feedback node. The red line corresponds to the fit of real data shown in Figure~\ref{fig:gbp}.}
%  \label{fig:gbp_mc}
%\end{figure}

Texas Instruments provides a SPICE model of the LMH6629 that was verified with Microcap 10 from Spectrum Software. While the noise simulation is accurate at better than a few percent within the temperature range of interest, the simulated bandwidth does not follow the experimental behavior of the amplifier, consistently predicting a smaller GBP than what was measured (up to a factor of more than 2) for temperatures below \SI{270}{\kelvin}.

\section{Transimpedance Amplifier}
\label{sec:tia}
Based on the performance of the LMH6629, a low-noise transimpedance amplifier was designed for cryogenic applications.
% and used to readout a large-area SiPM with single photon sensitivity and nanosecond timing resolution.

\begin{table}[!t]
\centering
\renewcommand{\arraystretch}{1.5}
\caption{$R_d$ and $C_d$ values for \DSkSiPMAreaMax\ \NUVHdLf\ \SiPMs\ at breakdown voltage.}
\begin{tabular}{lcccc}
\toprule
\multirow{2}{*}{Frequency} & $R^{300K}_d$ & $C^{300K}_d$ & $R^{77K}_d$ & $C^{77K}_d$ \\
& [\si{\ohm\square\cm}]& [\si{\nano\F\per\square\cm}] & [\si{\ohm\square\cm}] & [\si{\nano\F\per\square\cm}] \\
\midrule
\SI{10}{\kHz} & \num{35} & \num{6.3} & \num{74} & \num{6.7} \\
\SI{100}{\kHz} & \num{14} & \num{6.2} & \num{63} & \num{6.6} \\
\SI{200}{\kHz} & \num{13} & \num{6.2} & \num{63} & \num{6.5} \\
\SI{500}{\kHz} & \num{12} & \num{6.1} & \num{62} & \num{5.8} \\
\SI{1}{\MHz} & \num{11} & \num{5.9} & \num{61} & \num{4.2} \\
\bottomrule
\end{tabular}
\caption*{\footnotesize All measures have an uncertainty of 1 unit on the last digit.}
\label{tab:RCd}
\end{table}

\begin{figure}[!t]
  \centering
  \includegraphics[width=\columnwidth]{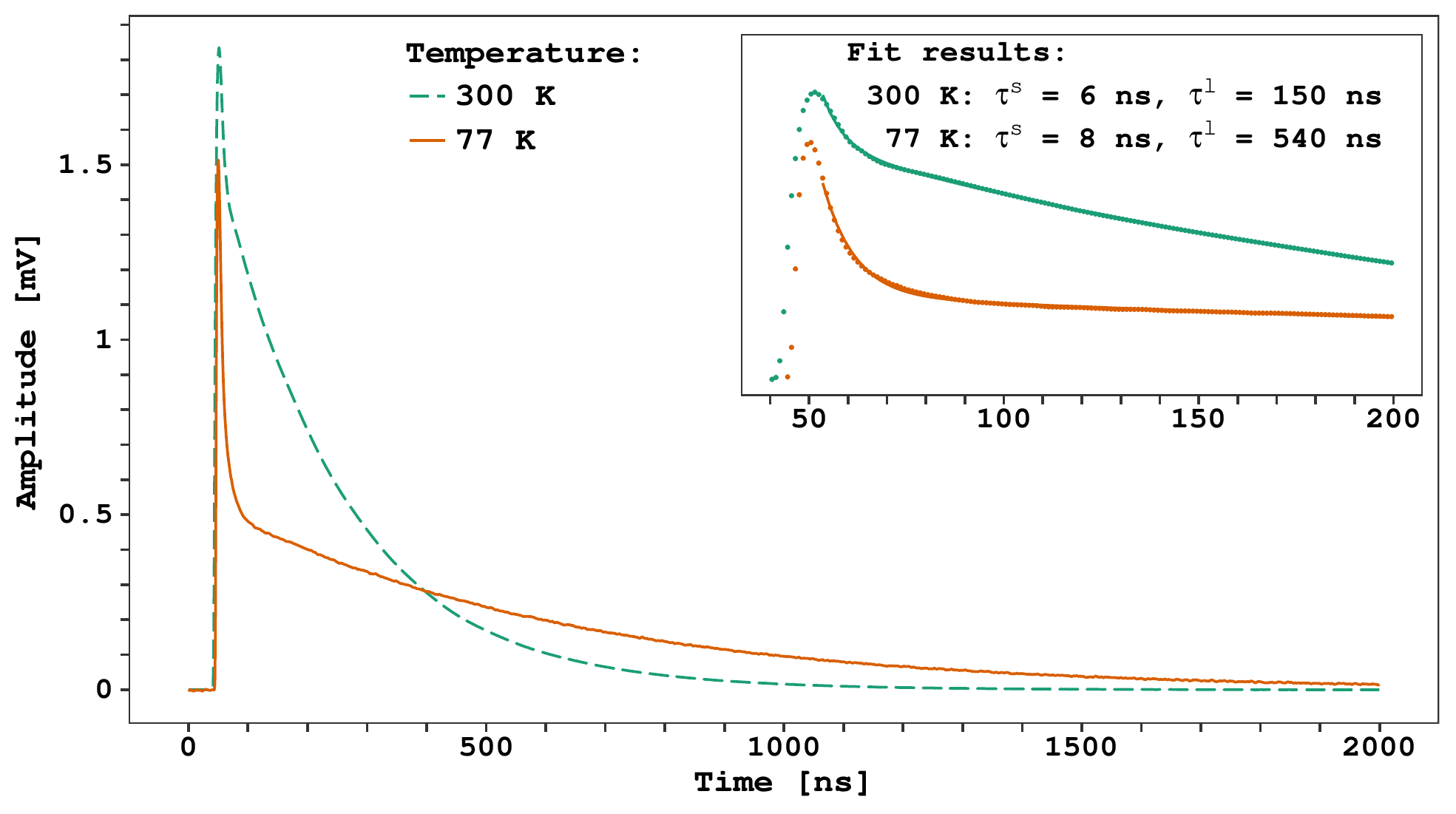}
  \caption{Average waveform of a \NUVHdLf\ \DSkSiPMAreaMax\ \SiPM\ photoelectron acquired with the TIA described in the text with $R_f = \SI{2}{\kohm}$ and $R_s = \SI{20}{\ohm}$. The inset shows a zoom of the fast peak (points) and the fit with a double exponential (solid lines); the rise time is about \SI{2}{\nano\second}, read-out limited.}
  \label{fig:shape}
\end{figure}

\subsection{\NUVHdLf\ \SiPM}
\label{sec:sipm}
The transimpedance amplifier was optimized for use with a custom batch of  \NUVHdLf\, \DSkSiPMAreaMax\ area \SiPMs\ produced by FBK for DarkSide-20k~\cite{Aalseth:2017tr}. These devices are similar to FBK's standard \SI{25}{\um} NUV-HD technology~\cite{Piemonte:2016cj} but are engineered to have a lower field strength in the avalanche region~\cite{Ferri:2016ky} and a smaller quenching resistance ($R_q= \SI{2.2}{\Mohm}$ at room temperature) to provide a relatively fast signal. The average photoelectron signal from a \NUVHdLf\ \DSkSiPMAreaMax\ \SiPM\ is shown in Figure~\ref{fig:shape}, both at room and liquid nitrogen temperature. These \SiPMs\ can be operated in liquid nitrogen with gains up to \SI{1.5E6}{\coulomb\per\coulomb}. In this work, the \SiPM\ operating voltage is tuned so that at each operating temperature, the gain remains stable at \SI{1E6}{\coulomb\per\coulomb}. %, allowing for a direct comparison of the noise measured in each configuration.

The recharge time constant of the \SiPM\ was determined over the interval \SIrange{60}{300}{\ns} by a fit to the formula 
\begin{equation}
  \tau^l = \tau_0 / T^\alpha
  \label{eqn:tau}
\end{equation}
where $\alpha=\num{1.04\pm0.01}$ and $\tau_0=\SI{51\pm3}{\micro\second\per\kelvin}$ with a standard deviation of the relative residuals of \SI{2}{\percent}.

In addition to the characteristics of the \NUVHdLf\ \SiPMs\ reported in~\cite{Acerbi:2017gy}, the detector equivalent capacitance ($C_d$) and its equivalent series resistance ($R_d$) are relevant for this work. As discussed in~\cite{Marano:2014}, these parameters are the result of the summing of all the microcell properties: resistance of the quenching resistor ($R_q$), capacitance of the quenching resistor ($C_q$) and the cell intrinsic capacitance ($C_\textrm{SPAD}$). The characterization of the microcell properties at cryogenic temperatures is beyond the scope of this paper. A direct measurement of $C_d$ and $R_d$ was done using an Agilent E4980A RLC meter and the results are reported in Table~\ref{tab:RCd}. %The behavior of $R_d$ is particularly relevant for the design of the pre-amplifier because the noise gain strongly depends on it.

Some observations can be made from the measurement of $C_d$ and $R_d$. First, while the detector capacitance remains relatively constant with temperature variation, $R_d$ increases in liquid nitrogen by about a factor of five. This is due to the large variation of $R_q$, which is a poly-silicon resistor. This in turn increases the SiPM recharge time. %Using the data at \SI{100}{\kilo\hertz} and Equation~\ref{eqn:tau}, $R_d$ can be modeled as function of temperature as 
    %\begin{equation}
      %R_d=R_0/T^\alpha
      %\label{eqn:rd}
    %\end{equation}
%where $R_0=\SI{5.4\pm0.3}{\kilo\ohm\per\kelvin}$. 
Second, both $R_d$ and $C_d$ vary as a function of frequency. To first order, $R_q$, $C_q$, and $C_\textrm{SPAD}$ form a filter network with one pole in $Fp = R_q\times C_q$, which at room temperature has a value in the range \SIrange[range-phrase = ~to~, range-units = single]{10}{20}{\mega\hertz} (see~\cite{Marano:2014}).  For higher frequencies, $R_d$ asymptotically approaches zero. At cryogenic temperature, the larger value of $R_q$ shifts the pole down to a few MHz. 
By assuming a ratio $C_\textrm{SPAD}$/$C_q \simeq \num{10}$, $Fp$ can be estimated as $\frac{1}{2\pi Rd\, Cd/10}$ corresponding to \SI{18}{\mega\hertz} at room temperature and \SI{4}{\mega\hertz} at \SI{77}{\kelvin}.

\subsection{Pre-amplifier design}
\begin{figure}[!t]
  \centering
  \includegraphics[width=0.92\columnwidth]{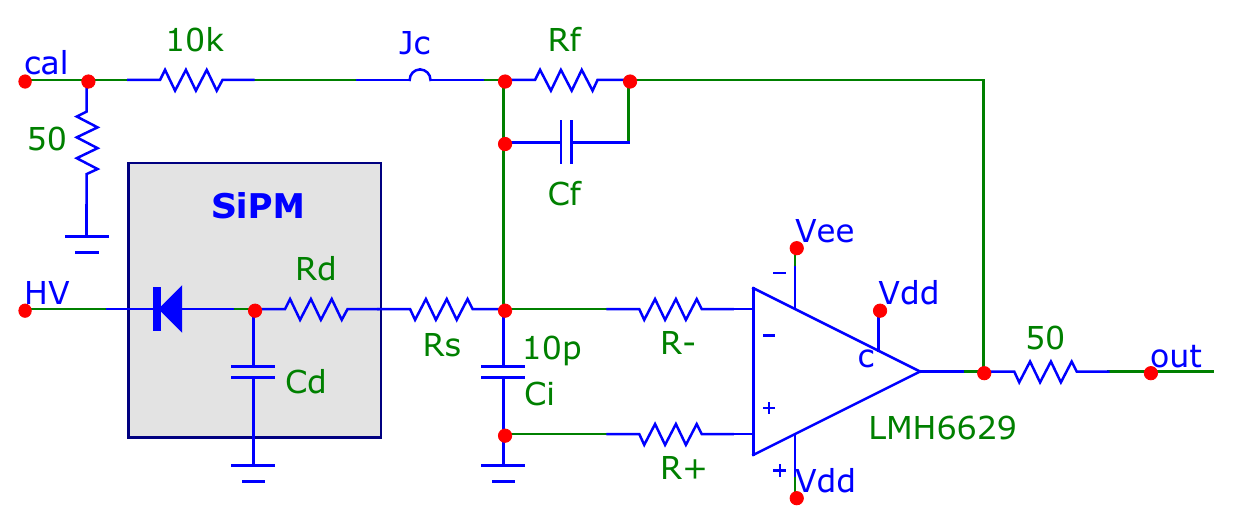}
  \caption{Schematic of the transimpedance amplifier described in the text. During regular operation the calibration input is disconnected (at $J_c$) to avoid injecting noise: the \SI{10}{\kilo\ohm} resistor ensures that the calibration input behaves as a current like source.}
  \label{fig:tia}
\end{figure}

The schematic of the transimpedance amplifier is shown in Figure~\ref{fig:tia}. The circuit is based on an LMH6629 operating with a \SI{4}{\volt} supply voltage and its on-chip compensation reduced. 

The resistors $R_-$ and $R_+$ (both \SI{10}{\ohm}) at the inverting and non-inverting inputs are required to avoid high frequency oscillations at cryogenic temperature. This is done at the cost of a roughly \SI{20}{\percent} increase in the voltage noise,  $e_n$. The \SI{10}{\pF} capacitor $C_i$ minimizes undershoot due to the fast discharge of the SiPM. The feedback capacitance, $C_f$, is only a few hundred \si{\femto\farad} and is due to the stray capacitance of the printed circuit board, as discussed in Section~\ref{sec:bandwidth}. 

\begin{figure}[!t]
  \centering
  \includegraphics[width=\columnwidth]{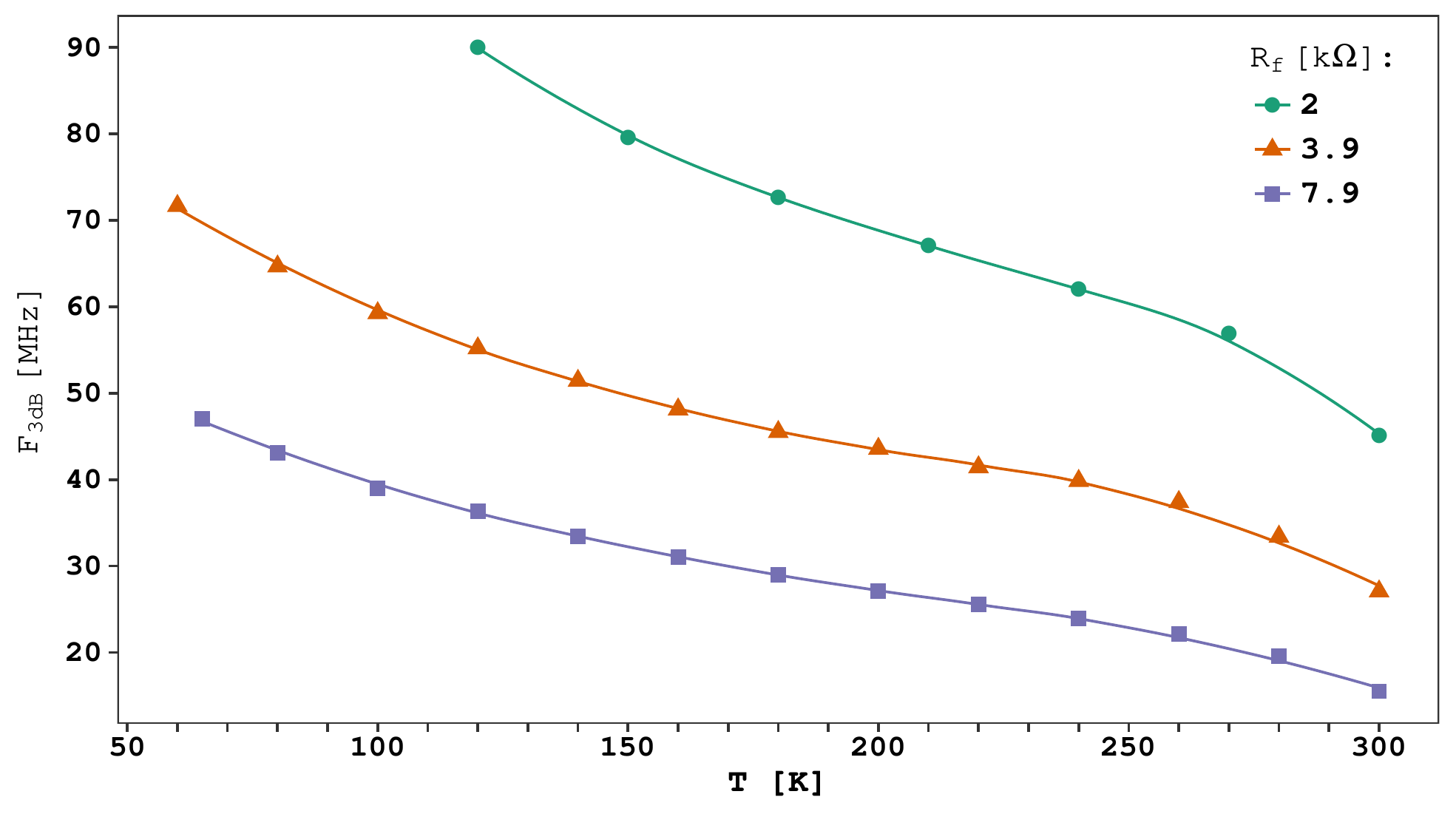}
  \caption{$F_\textrm{-3dB}$ of the transimpedance amplifier discussed in the text. The lines are drawn to guide the eye.}
  \label{fig:tia_bw}
%  \vskip 3mm
\end{figure}

The series resistor, $R_s$ (and to some extent $R_d$), plays an important role in the performance of the amplifier. For frequencies above $F_l = \frac{1}{2 \pi C_d \left( R_s + R_d \right)}$, the system behaves like an inverting amplifier due to the presence of $R_s$ and $R_d$. This means that the bandwidth is proportional to the GBP of the amplifier, in contrast to an ideal TIA, where the bandwidth scales as the square root of the GBP. The result is an increased bandwidth. Figure~\ref{fig:tia_bw} shows the temperature scaling of the pre-amplifier bandwidth as measured by $S_{21}$ VNA scans between the calibration input and the output with the SiPM connected. In addition, the series resistor changes the noise gain so that, for frequencies above $F_l$, it plateaus at $N_g = \frac{R_f}{R_s + R_d} + 1$. This in turn reduces the output noise without changing the transfer function for current sources (like \SiPMs).

However, large values of $R_s$ will alter the output signal shape by creating a low-pass filter with the detector capacitance and, more importantly, significantly contribute to the noise budget.

\subsection{Pre-amplifier performance in liquid nitrogen}
\label{sec:pre:performances}
\begin{figure}[!t]
  \centering
  \includegraphics[width=\columnwidth]{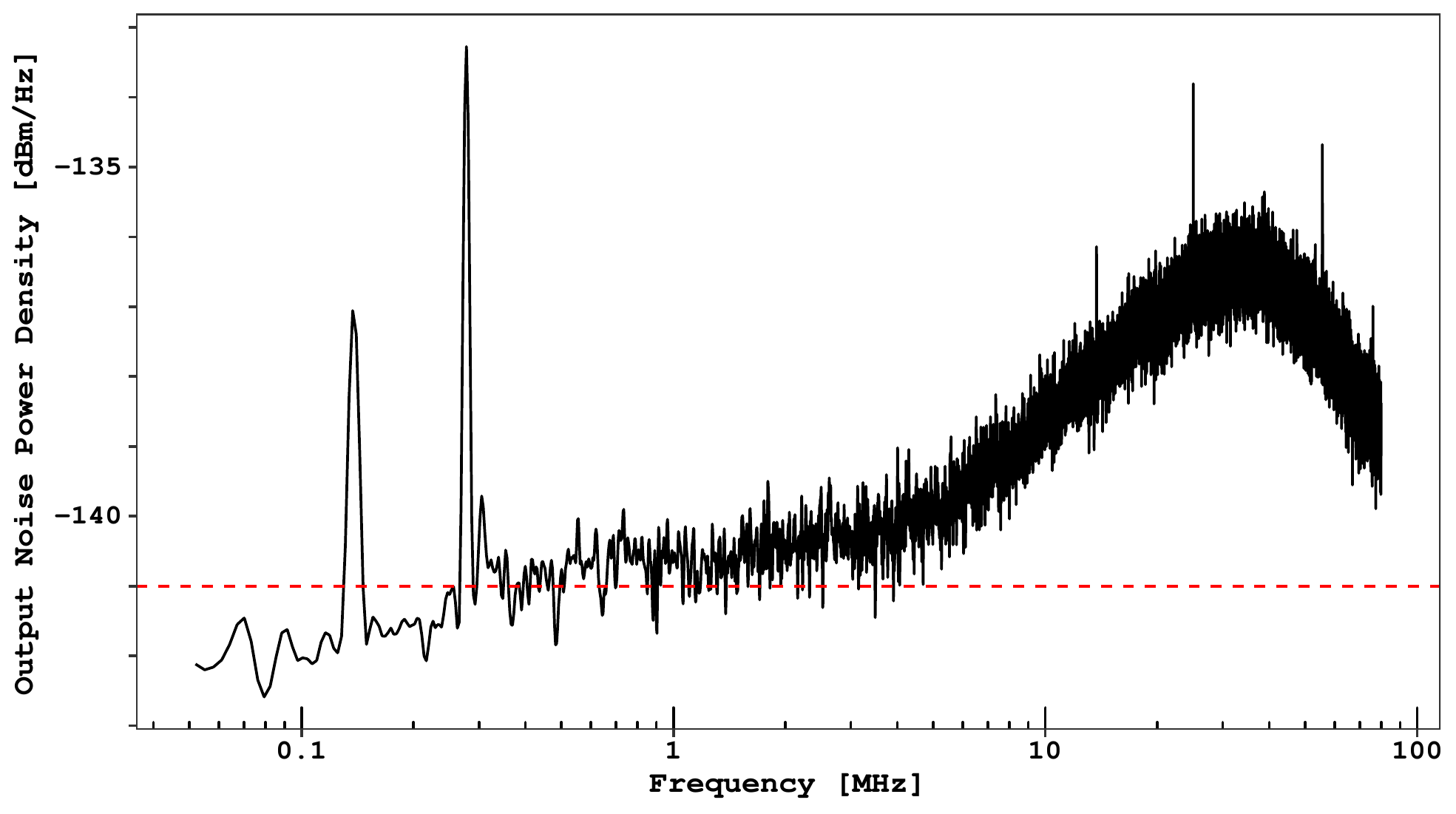}
  \caption{Output noise density of the pre-amplifier discussed in the text at \SI{77}{\kelvin} with the \SiPM\ connected. The red line shows the analytical noise prediction.}
  \label{fig:noise_density}
\end{figure}

Values of $R_f = \SI{3.9}{\kohm}$ and $R_s = \SI{20}{\ohm}$ were selected to optimize the TIA performance at \SI{77}{\kelvin}. With these values, the bandwidth of the pre-amplifier is always significantly higher than the fast rising edge of the \SiPM\ (which is between  
\SIrange[range-phrase = ~and~, range-units = single]{20}{30}{\MHz} and is extracted from the fit of the fast component of the waveform shown in Figure~\ref{fig:shape}).

Above $F_l$, the voltage noise at the input of the amplifier, $e^t_n(T) = \sqrt{4 k_B T (R_n + R_+ + R_- + R_s + R_d)}$, includes the intrinsic voltage noise of the LMH6629, $e_n$, and the Johnson-Nyquist noise of the other resistors in the circuit. In this formula, $R_n = \SI{25}{\ohm}$ accounts for $e_n$, as described in Section~\ref{sec:noise}. In liquid nitrogen  $e^t_n(\SI{77}{\kelvin}) = \SI{0.7}{\nano\volt\per\sqrt\hertz}$ corresponding to about \SI{-141}{dBm\per\hertz} at the output (taking into account the back termination of the output).
The input current noise of the LMH6629, $i_n$, and the Johnson-Nyquist current noise of $R_f$ together account for about \SI{1.5}{\pico\ampere\per\sqrt\hertz}, which corresponds to about \SI{-157}{dBm\per\hertz} at the output and can therefore be ignored.
Figure~\ref{fig:noise_density} shows the output noise density spectrum of the pre-amplifier connected to the \NUVHdLf\ \DSkSiPMAreaMax\ \SiPM. The difference at \SI{1}{\MHz} between the prediction and the measurement is less than \SI{1}{dB}. 

The high frequency bump present in the noise spectrum is due to the decrease of $R_d$ at few MHz as described at the end of Section~\ref{sec:sipm}. With a vanishing $R_d$, the noise gain increases up to a theoretical factor of \num{4}. In practice this value is never reached due to the limited available GBP.

With a \SI{4}{\volt} power supply, the pre-amplifier provides a dynamic range in excess of \num{250} photo-electrons with a power consumption of about \SI{40}{\milli\watt}. % (see Section~\ref{sec:dyn} and~\ref{sec:power}).

%At room temperature the output noise is increases to about \SI{500}{\micro\volt}. At \SI{300}{\kelvin} $e^t_n$ raises by a factor \num{1.5} as dictated by the Johnson-Nyquist scaling. Moreover the noise gain increases by a factor 2.6 thanks to the smaller value of $R_d$.

\subsection{Readout and Data analysis}
\label{sec:analysis}
\begin{figure}[!t]
  \centering
  \includegraphics[width=\columnwidth]{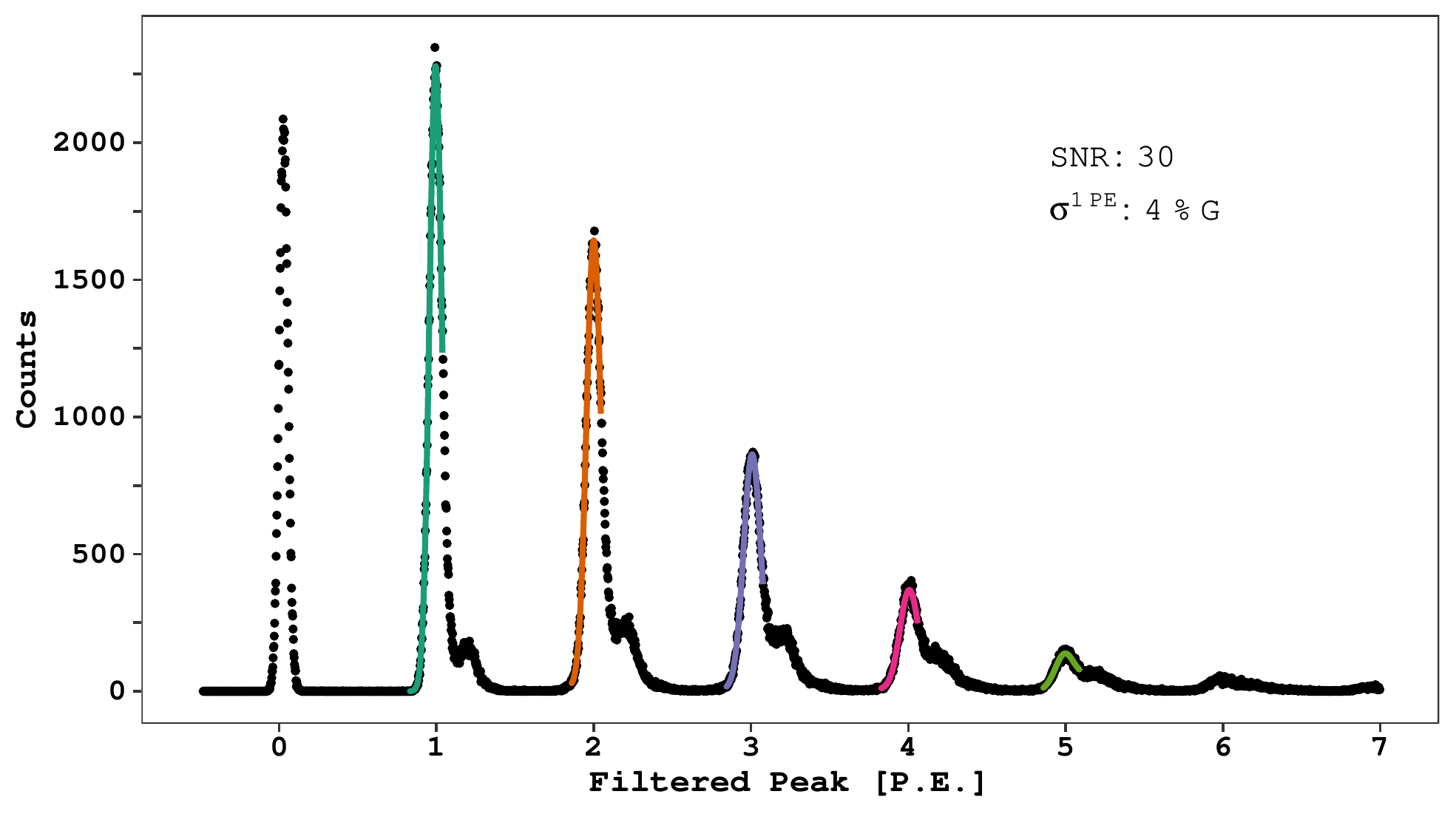}
  \caption{Spectrum of photoelectrons after matched filtering for a \DSkSiPMAreaMax\ \NUVHdLf\ \SiPM\ at \SI{77}{\kelvin}}
  \label{fig:fingers}
\end{figure}

Cryogenic tests of the transimpedance amplifier with the \DSkSiPMAreaMax\ \NUVHdLf\ \SiPM\  were performed in the setup described in Section~\ref{sec:intro}. A Hamamatsu PLP-10 \SI{405}{\nano\meter} laser, capable of sub-nanosecond light pulses, was used as a light source. The signal from the transimpedance amplifier was fed out of the cryostat into a room temperature, low-noise amplifier and then digitized by a 10 bit, 2 GS/s \LNGSCryoSetupDigitizerModel\ configured with a pre-trigger of \SI{8}{\micro\s} and a total gate length of \SI{15}{\micro\s}. The digitizer was triggered by the sync pulse of the PLP-10. The least significant bit of the digitizer is \SI{1}{mV} so further amplification of the signal is required to avoid excessive quantization noise. A custom low noise amplifier based on two THS3201 operational amplifiers~\cite{THS3201} was built providing a bandwidth in excess of \SI{200}{MHz}, a gain of \SI{100}{\volt\per\volt}, and an input noise density of about \SI{1.6}{\nano\volt\per\sqrt\hertz}.

Offline analysis software was used to analyze the digitized waveforms and calculate the figures of merit. Of particular interest is the signal to noise ratio (SNR), defined as the ratio between the gain of the photoelectron peaks and the baseline noise. Two SNR parameters were calculated based on two different methods for calculating the charge of each waveform: the amplitude of the raw waveform at the peak ($\mathbf{SNR_{r}}$) and the amplitude of the digitally filtered waveform at the peak ($\mathbf{SNR_{f}}$).
In both cases, the baseline noise was estimated as the standard deviation of the corresponding pre-trigger raw or filtered waveform. 

\subsection{Matched filter}
\label{sec:filter}
The extraction of a signal with known shape from stochastic noise is maximized by the use of a matched filter~\cite{Turin:1960kw}. Matched filtering is done by convolving the raw waveform with a time reversed template waveform. The matched filter algorithm is implemented offline. For a given pre-amplifier configuration and temperature, the the template waveform is determined by averaging baseline subtracted and time aligned waveforms from single photoelectron events. The filtered waveform exhibits a symmetric shape around the photo-electron arrival time, where a peak is present. The identification of the charge and timing of the signal is performed by evaluating the amplitude and the timing of the filtered waveform maximum. %Based on the symmetric nature of the calculated peaks, it is clear that the fluctuation of these estimators is dominated by statistical variation (i.e. the contribution of random noise does not bias the central value).

\begin{figure}
    %  \vskip 3mm
  \includegraphics[width=\columnwidth]{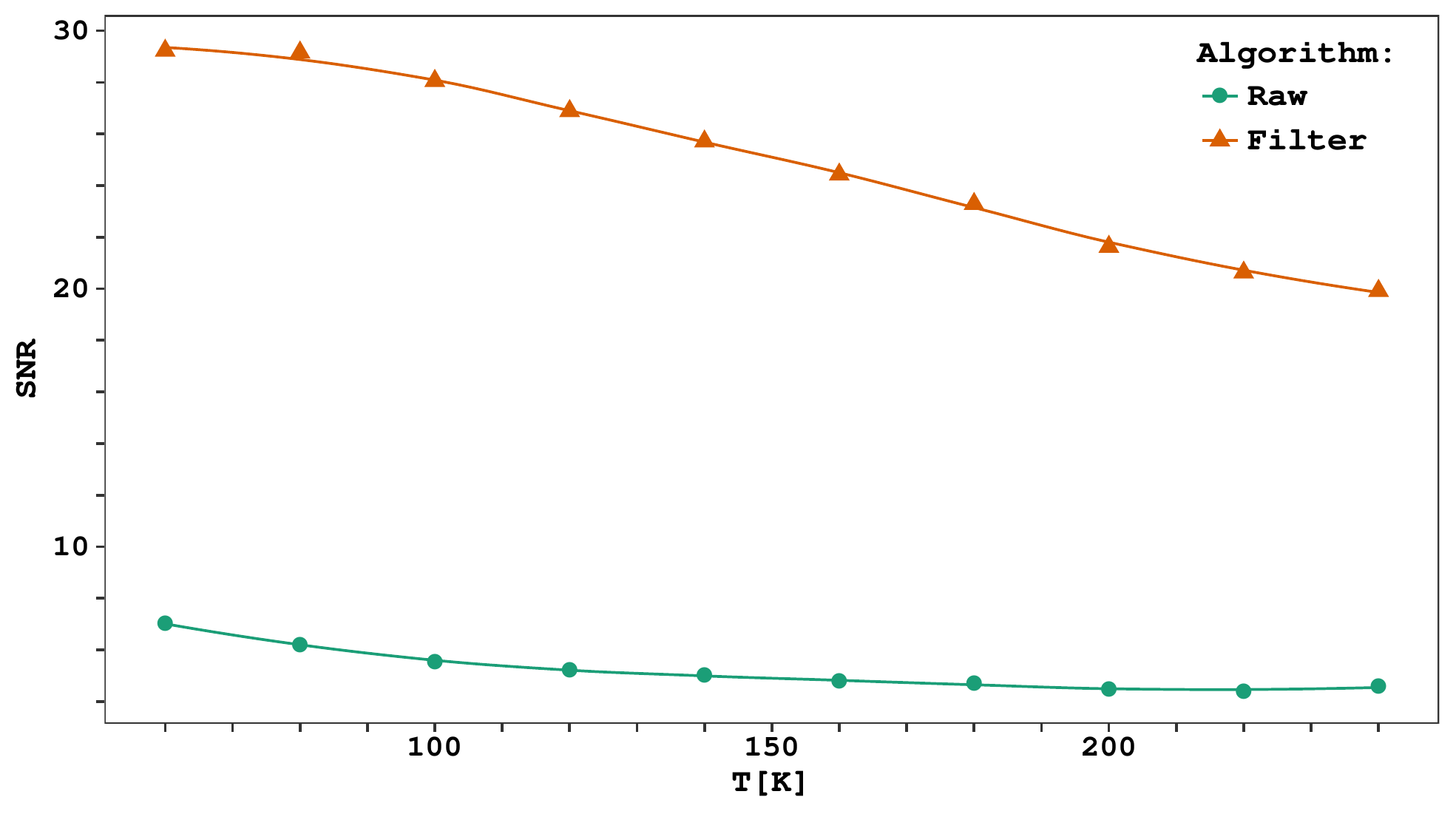}
  \caption{Signal to noise figures as a function of temperature. The lines are drawn to guide the eye.}
  \label{fig:SNR}
\end{figure}

\section{Results}
\label{sec:results}
The spectrum obtained using a \DSkSiPMAreaMax\ \NUVHdLf\ \SiPM\ at \SI{77}{\kelvin} and the matched filter algorithm described in Section~\ref{sec:filter} is shown in Figure~\ref{fig:fingers}. An SNR of \num{30} is achieved with the \SiPM\ operating at a gain of \SI{1E6}{\coulomb\per\coulomb}.

\subsection{Signal over noise}
Figure~\ref{fig:SNR} compares $\mathbf{SNR_{f}}$ and $\mathbf{SNR_{r}}$ as a function of temperature. The behavior of these curves includes many effects. To first order, both the signal amplitude and the baseline noise increase with higher temperature. A simple model that only considers the temperature dependence of the amplitude and the noise plateau over-predicts the decrease in SNR at room temperature. This is because the magnitude and shape of the output noise spectrum changes due to the temperature dependence of the \TIA\ bandwidth, $F_l$, and $F_p$. The net effect on the baseline noise at the bandwidth of interest, which is also temperature dependent,  is therefore difficult to model analytically. %We believe that a full analytical model would not add anything to this work and therefore is not further investigated.

\subsection{Timing}
\begin{figure}[!t]
  \centering
  \includegraphics[width=\columnwidth]{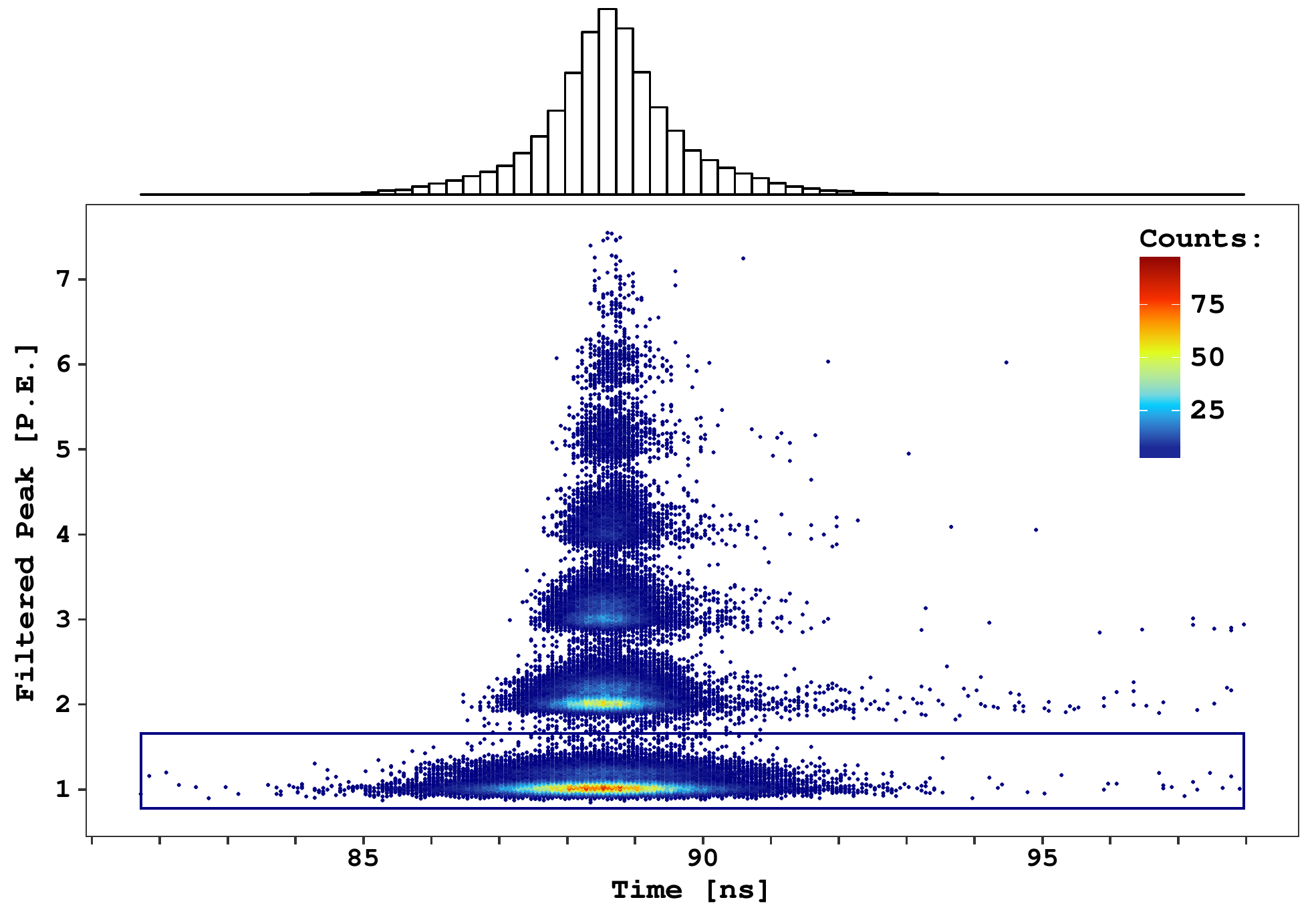}
  \caption{Peak amplitude of the matched filtered waveform versus the reconstructed time relative to the laser sync at \SI{80}{\kelvin}. The top histogram shows the timing distribution of the single photoelectron peak (within the black box). The standard deviation of the time is \SI{1.0}{\nano\second}.}
  \label{fig:timespread}
\end{figure}

The rise-time of the unfiltered \SiPM\ signal is a function of the amplifier bandwidth shown in Figure~\ref{fig:tia_bw} and is between 5 and 11~ns (modeling the rise-time as 0.35/bandwidth). Given the relatively low $\mathbf{SNR_{r}}$, a standard discriminator is not the best approach for determining the event timing. In situations like this, it is advantageous to use the entire \SiPM\ waveform for event reconstruction. This is particularly true at cryogenic temperature, when a large fraction of the total signal charge is contained in the slow recharge of the waveform. This is done using matched filtering, which preserves the fast component of the signal while maximizing the SNR and yielding nanosecond timing resolution. Figure~\ref{fig:timespread} shows the time jitter of the reconstructed event time relative to the laser pulse measured using the matched filtered waveform at \SI{80}{\kelvin}. The standard deviation of the first photoelectron peak is \SI{1.0}{\nano\second}. 

\begin{figure}[!t]
  \centering
  %  \vskip 3mm
  \includegraphics[width=\columnwidth]{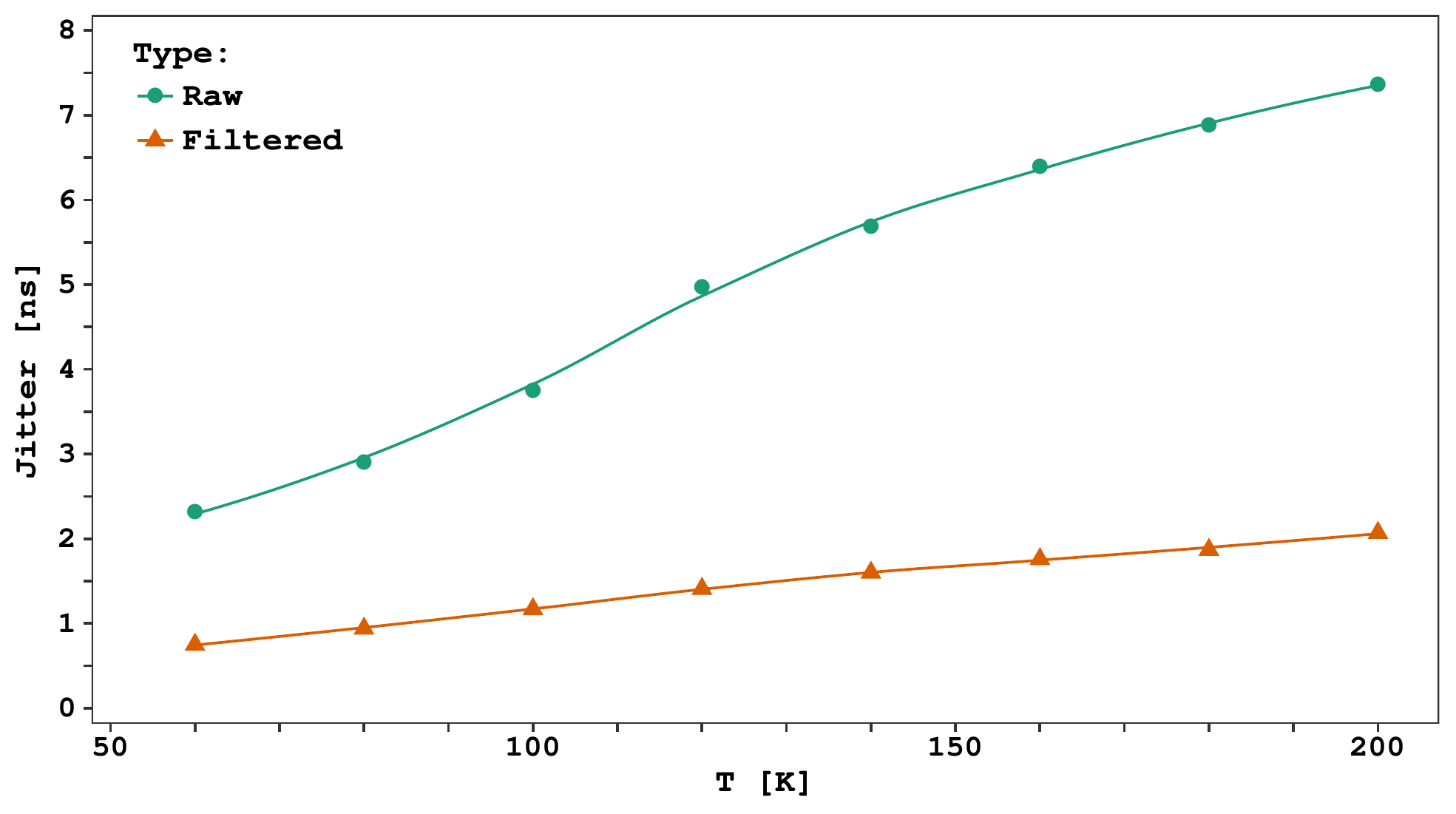}
  \caption{Standard deviation of the calculated event time as a function of temperature for both matched filtered and raw waveforms. For the latter, a simple discriminator was used to determine the time of the event. The lines are drawn to guide the eye.}
  \label{fig:tres_vs_t}
\end{figure}

Figure~\ref{fig:tres_vs_t} shows the time distribution as a function of temperature for both filtered and unfiltered waveforms (in this case using a simple threshold to define the timing). While the timing resolution of the filtered waveform is slightly affected by the temperature, the performance of the unfiltered waveform shows a strong degradation with temperature. This is expected because the jitter of a simple discrimination can be quantified as the quotient of the risetime divided by the signal to noise ratio. In our setup the SNR decreases with temperature (Figure~\ref{fig:SNR}) and the bandwidth decreases with temperature (Figure~\ref{fig:tia_bw}), significantly reducing the fast rising edge of the waveform.

\section{Conclusions}

SiPM-based, cryogenic photo-detectors with large active areas have great appeal for use in low-background particle physics experiments, where improvements in photo-detection efficiency, resolution, and radio-purity directly impact experimental reach. The viability of such photo-detectors relies on the development of cryogenic pre-amplifiers capable of preserving the fidelity of photon signals despite large sensor capacitance. In this paper, measurements of the electrical characteristics of a Texas Instruments LMH6629 were made over a range of temperatures from room temperature to \SI{60}{\kelvin} and a cryogenic transimpedance amplifier was designed around the device. The resulting amplifier design was operated with a \NUVHdLf\, \DSkSiPMAreaMax\ \SiPM\ with an SNR of \num{45} at \SI{77}{\kelvin} when operating the \SiPM\ at a gain of \SI{1.5E6}{\coulomb\per\coulomb} (\SI{5}{\volt} of overvoltage) and timing resolution better than \SI{1}{\nano\second}, thus demonstrating the practicality of large-area SiPM detectors as conventional photomultiplier tube replacements for cryogenic applications.
 
\label{sec:conclusions}

\bibliographystyle{IEEEtran}
\bibliography{$TEXMFHOME/ds,cryo-pre}

\begin{thebibliography}{10}
\providecommand{\url}[1]{#1}
\csname url@samestyle\endcsname
\providecommand{\newblock}{\relax}
\providecommand{\bibinfo}[2]{#2}
\providecommand{\BIBentrySTDinterwordspacing}{\spaceskip=0pt\relax}
\providecommand{\BIBentryALTinterwordstretchfactor}{4}
\providecommand{\BIBentryALTinterwordspacing}{\spaceskip=\fontdimen2\font plus
\BIBentryALTinterwordstretchfactor\fontdimen3\font minus
  \fontdimen4\font\relax}
\providecommand{\BIBforeignlanguage}[2]{{%
\expandafter\ifx\csname l@#1\endcsname\relax
\typeout{** WARNING: IEEEtran.bst: No hyphenation pattern has been}%
\typeout{** loaded for the language `#1'. Using the pattern for}%
\typeout{** the default language instead.}%
\else
\language=\csname l@#1\endcsname
\fi
#2}}
\providecommand{\BIBdecl}{\relax}
\BIBdecl

\bibitem{Acerbi:2017gy}
\BIBentryALTinterwordspacing
F.~Acerbi, S.~Davini, A.~Ferri, C.~Galbiati, G.~Giovanetti, A.~Gola, G.~Korga,
  A.~Mandarano, M.~Marcante, G.~Paternoster, C.~Piemonte, A.~Razeto,
  V.~Regazzoni, D.~Sablone, C.~Savarese, G.~Zappala, and N.~Zorzi, ``{Cryogenic
  Characterization of FBK HD Near-UV Sensitive SiPMs},'' \emph{IEEE Trans.
  Elec. Dev.}, pp. 1--6, 2017. [Online]. Available:
  \url{http://ieeexplore.ieee.org/document/7807295/}
\BIBentrySTDinterwordspacing

\bibitem{Aalseth:2017tr}
\BIBentryALTinterwordspacing
C.~E. Aalseth, F.~Acerbi, P.~Agnes, I.~F.~M. Albuquerque, T.~Alexander,
  A.~Alici, A.~K. Alton, P.~Antonioli, S.~Arcelli, R.~Ardito, I.~J. Arnquist,
  D.~M. Asner, M.~Ave, H.~O. Back, A.~I.~B. Olmedo, G.~Batignani, E.~Bertoldo,
  S.~Bettarini, M.~G. Bisogni, V.~Bocci, A.~Bondar, G.~Bonfini, W.~Bonivento,
  M.~Bossa, B.~Bottino, M.~G. Boulay, R.~Bunker, S.~Bussino, A.~F. Buzulutskov,
  M.~Cadeddu, M.~Cadoni, A.~Caminata, N.~Canci, A.~Candela, C.~Cantini,
  M.~Caravati, M.~Cariello, M.~Carlini, M.~Carpinelli, A.~Castellani,
  S.~Catalanotti, V.~Cataudella, P.~Cavalcante, S.~Cavuoti, R.~Cereseto,
  A.~Chepurnov, C.~Cical{\`o}, L.~Cifarelli, M.~Citterio, A.~G. Cocco,
  M.~Colocci, S.~Corgiolu, G.~Covone, P.~Crivelli, I.~D'Antone, M.~D'Incecco,
  D.~D{\textquoteright}Urso, M.~D. Da~Rocha~Rolo, M.~Daniel, S.~Davini,
  A.~De~Candia, S.~De~Cecco, M.~De~Deo, G.~De~Filippis, G.~De~Guido,
  G.~De~Rosa, G.~Dellacasa, M.~Della~Valle, P.~Demontis, A.~Derbin, A.~Devoto,
  F.~Di~Eusanio, G.~Di~Pietro, C.~Dionisi, A.~Dolgov, I.~Dormia, S.~Dussoni,
  A.~Empl, M.~F. Diaz, A.~Ferri, C.~Filip, G.~Fiorillo, K.~Fomenko, D.~Franco,
  G.~E. Froudakis, F.~Gabriele, A.~Gabrieli, C.~Galbiati, P.~G. Abia,
  A.~Gendotti, A.~Ghisi, S.~Giagu, P.~Giampa, G.~Gibertoni, C.~Giganti, M.~A.
  Giorgi, G.~K. Giovanetti, M.~L. Gligan, A.~Gola, O.~Gorchakov, A.~M. Goretti,
  F.~Granato, M.~Grassi, J.~W. Grate, G.~Y. Grigoriev, M.~Gromov, M.~Guan,
  M.~B.~B. Guerra, M.~Guerzoni, M.~Gulino, R.~K. Haaland, A.~Hallin, B.~Harrop,
  E.~W. Hoppe, S.~Horikawa, B.~Hosseini, D.~Hughes, P.~Humble, E.~V.
  Hungerford, A.~M. Ianni, C.~Jillings, T.~N. Johnson, K.~Keeter, C.~L.
  Kendziora, S.~Kim, G.~Koh, D.~Korablev, G.~Korga, A.~Kubankin, M.~Kuss,
  B.~Lehnert, X.~Li, M.~Lissia, G.~U. Lodi, B.~Loer, G.~Longo, P.~Loverre,
  R.~Lussana, L.~Luzzi, Y.~Ma, A.~A. Machado, I.~N. Machulin, A.~Mandarano,
  L.~Mapelli, M.~Marcante, A.~Margotti, S.~M. Mari, M.~Mariani, J.~Maricic,
  C.~J. Martoff, M.~Mascia, M.~Mayer, A.~B. McDonald, A.~Messina, P.~D. Meyers,
  R.~Milincic, A.~Moggi, S.~Moioli, J.~Monroe, A.~Monte, M.~Morrocchi, B.~J.
  Mount, W.~Mu, V.~N. Muratova, S.~Murphy, P.~Musico, R.~Nania, A.~N. Agasson,
  I.~Nikulin, V.~Nosov, A.~O. Nozdrina, N.~N. Nurakhov, A.~Oleinik,
  V.~Oleynikov, M.~Orsini, F.~Ortica, L.~Pagani, M.~Pallavicini, S.~Palmas,
  L.~Pandola, E.~Pantic, E.~Paoloni, G.~Paternoster, V.~Pavletcov, F.~Pazzona,
  S.~Peeters, K.~Pelczar, L.~A. Pellegrini, N.~Pelliccia, F.~Perotti,
  R.~Perruzza, V.~P. Fortes, C.~Piemonte, F.~Pilo, A.~Pocar, T.~Pollmann,
  D.~Portaluppi, D.~A. Pugachev, H.~Qian, B.~Radics, F.~Raffaelli, F.~Ragusa,
  M.~Razeti, A.~Razeto, V.~Regazzoni, C.~Regenfus, B.~Reinhold, A.~L. Renshaw,
  M.~Rescigno, F.~Retiere, Q.~Riffard, A.~Rivetti, S.~Rizzardini, A.~Romani,
  L.~Romero, B.~Rossi, N.~Rossi, A.~Rubbia, D.~Sablone, P.~Salatino,
  O.~Samoylov, E.~S. Garc{\'\i}a, W.~Sands, M.~Sant, R.~Santorelli,
  C.~Savarese, E.~Scapparone, B.~Schlitzer, G.~Scioli, E.~Segreto, A.~Seifert,
  D.~A. Semenov, A.~Shchagin, L.~Shekhtman, E.~Shemyakina, A.~Sheshukov,
  M.~Simeone, P.~N. Singh, P.~Skensved, M.~D. Skorokhvatov, O.~Smirnov,
  G.~Sobrero, A.~Sokolov, A.~Sotnikov, F.~Speziale, R.~Stainforth, C.~Stanford,
  G.~B. Suffritti, Y.~Suvorov, R.~Tartaglia, G.~Testera, A.~Tonazzo, A.~Tosi,
  P.~Trinchese, E.~V. Unzhakov, A.~Vacca, E.~V{\'a}zquez-J{\'a}uregui,
  M.~Verducci, T.~Viant, F.~Villa, A.~Vishneva, R.~B. Vogelaar, M.~Wada,
  J.~Wahl, J.~Walding, S.~E. Walker, H.~Wang, Y.~Wang, A.~W. Watson,
  S.~Westerdale, R.~Williams, M.~M. Wojcik, S.~Wu, X.~Xiang, X.~Xiao, C.~Yang,
  Z.~Ye, A.~Y. de~Llano, F.~Zappa, G.~Zappal{\`a}, C.~Zhu, A.~Zichichi,
  M.~Zullo, and A.~Zullo, ``{DarkSide-20: A 20 Tonne Two-Phase LAr TPC for
  Direct Dark Matter Detection at LNGS},'' \emph{arXiv}, 2017. [Online].
  Available: \url{http://arxiv.org/abs/1707.08145v1}
\BIBentrySTDinterwordspacing

\bibitem{Dumke:1981dy}
\BIBentryALTinterwordspacing
W.~P. Dumke, ``{The effect of base doping on the performance of Si bipolar
  transistors at low temperatures},'' \emph{IEEE Trans. Elec. Dev.}, vol.~28,
  no.~5, pp. 494--500, 1981. [Online]. Available:
  \url{http://ieeexplore.ieee.org/document/1481524/}
\BIBentrySTDinterwordspacing

\bibitem{Cressler:1994ck}
\BIBentryALTinterwordspacing
J.~D. Cressler, ``{Operation of SiGe bipolar technology at cryogenic
  temperatures},'' \emph{J. Phys. IV France}, vol.~04, no.~C6, pp.
  C6--101--C6--110, 1994. [Online]. Available:
  \url{http://www.edpsciences.org/10.1051/jp4:1994616}
\BIBentrySTDinterwordspacing

\bibitem{TexasInstruments:2016wl}
\BIBentryALTinterwordspacing
{Texas Instruments}, ``{LMH6629 Ultra-Low Noise, High-Speed Operational
  Amplifier with Shutdown},'' Texas Instruments, Tech. Rep., 2016. [Online].
  Available: \url{www.ti.com/lit/ds/symlink/lmh6629.pdf}
\BIBentrySTDinterwordspacing

\bibitem{Teyssandier:2010tw}
\BIBentryALTinterwordspacing
F.~Teyssandier and D.~Pr{\^e}le, ``{Commercially Available Capacitors at
  Cryogenic Temperatures},'' in \emph{WOLTE9}, 2010. [Online]. Available:
  \url{https://hal.archives-ouvertes.fr/hal-00623399}
\BIBentrySTDinterwordspacing

\bibitem{Piemonte:2016cj}
\BIBentryALTinterwordspacing
C.~Piemonte, F.~Acerbi, A.~Ferri, A.~Gola, G.~Paternoster, V.~Regazzoni,
  G.~Zappala, and N.~Zorzi, ``{Performance of NUV-HD Silicon Photomultiplier
  Technology},'' \emph{IEEE Trans. Elec. Dev.}, vol.~63, no.~3, pp. 1111--1116,
  2016. [Online]. Available:
  \url{http://ieeexplore.ieee.org/lpdocs/epic03/wrapper.htm?arnumber=7397984}
\BIBentrySTDinterwordspacing

\bibitem{Ferri:2016ky}
\BIBentryALTinterwordspacing
A.~Ferri, F.~Acerbi, A.~Gola, G.~Paternoster, C.~Piemonte, and N.~Zorzi,
  ``{Performance of FBK low-afterpulse NUV silicon photomultipliers for PET
  application},'' \emph{JINST}, vol.~11, no.~03, pp. P03\,023--P03\,023, 2016.
  [Online]. Available:
  \url{http://stacks.iop.org/1748-0221/11/i=03/a=P03023?key=crossref.3b90668de68c0dbe8d723d4154a2ece8}
\BIBentrySTDinterwordspacing

\bibitem{Marano:2014}
\BIBentryALTinterwordspacing
D.~Marano, M.~Belluso, G.~Bonanno, S.~Billotta, A.~Grillo, S.~Garozzo,
  G.~Romeo, O.~Catalano, G.~L. Rosa, G.~Sottile, D.~Impiombato, and
  S.~Giarrusso, ``Silicon photomultipliers electrical model extensive
  analytical analysis,'' \emph{IEEE Transactions on Nuclear Science}, vol.~61,
  no.~1, pp. 23--34, Feb 2014. [Online]. Available:
  \url{http://ieeexplore.ieee.org/document/6661445/}
\BIBentrySTDinterwordspacing

\bibitem{THS3201}
\BIBentryALTinterwordspacing
{Texas Instruments}, ``{1.8-GHz, Low Distortion, Current-Feedback Amplifier},''
  Texas Instruments, Tech. Rep., 2003. [Online]. Available:
  \url{http://www.ti.com/lit/ds/slos416c/slos416c.pdf}
\BIBentrySTDinterwordspacing

\bibitem{Turin:1960kw}
\BIBentryALTinterwordspacing
G.~Turin, ``{An introduction to matched filters},'' \emph{IEEE Trans. Inform.
  Theory}, vol.~6, no.~3, pp. 311--329, 1960. [Online]. Available:
  \url{http://ieeexplore.ieee.org/document/1057571/}
\BIBentrySTDinterwordspacing

\end{thebibliography}
\end{document}